\documentclass[prb,aps, reprint, twocolumn, groupedaddress, showpacs]{revtex4-1}

\usepackage{bm}
\usepackage{graphicx}
\usepackage{color}
\usepackage{gensymb}
\usepackage{amsmath}
\usepackage{amssymb}

\begin{document}

\title{Tunable two-dimensional electron gas at the surface of thermoelectric material In$_4$Se$_3$}

\author{K. Fukutani,$^{1,2}$ T. Sato,$^1$ P. V. Galiy,$^3$ K. Sugawara,$^4$ and T. Takahashi$^{1,4}$}

\affiliation{$^1$Department of Physics, Tohoku University, Sendai 980-8578, Japan\\
$^2$Institute for Excellence in Higher Education, Tohoku University, Sendai 980-8576, Japan\\
$^3$Electronics Department, I. Franko Lviv National University, 79005 Lviv, Ukraine\\
$^4$WPI Research Center, Advanced Institute for Materials Research, Tohoku University, Sendai 980-8577, Japan
}

\date{\today}

\begin{abstract}
We report the discovery of two-dimensional electron gas (2DEG) at the surface of thermoelectric material In$_4$Se$_3$ by angle-resolved photoemission spectroscopy. The observed 2DEG exhibits a nearly isotropic band dispersion with a considerably small effective mass of \textit{m}$^{*}$ = 0.16 \textit{m}$_0$, and its carrier density shows a significant temperature dependence, leading to unconventional metal-semiconductor transition at the surface. The observed wide-range thermal tunability of 2DEG in In$_4$Se$_3$ gives rise to additional degrees of freedom to better control the surface carriers of semiconductors. 
\end{abstract}


\maketitle
\section{INTRODUCTION}
The two-dimensional electron gas (2DEG) is an accumulation of electrons usually found at interfaces of semiconductors and insulators \cite{Interface2DEG, STO_LAO}, in which electrons are free to move within the plane of interface, while strongly localized in the direction perpendicular to it. The properties of 2DEGs are known to be sensitive to external stimuli and can be tuned by various means, such as electric and/or magnetic fields. The decades of studies on 2DEGs have not only led to the discoveries of various fascinating fundamental phenomena such as quantum Hall effects \cite{Klitzing, Tsui} and two-dimensional superconductivity \cite{2Dsuper}, but also played a pivotal role in the technological advancements like field-effect transistors in which the tunable 2DEG carriers play an essential role \cite{RMP_Ando}. Furthermore, two-dimensional confinement of electrons at interfaces has proven successful in improving the thermoelectric properties of materials\cite{Dresselhaus_PbTe}. The numerous studies \cite{SiGe, Bi2Te3, Harman, ReviewTE, Ohta_ZT, Ohta_Seebeck}, which were devoted to control such systems, have in fact greatly advanced the field of thermoeletrics.

2DEGs have been also discovered at surfaces of semiconductors, which are formed as a result of strong shift of energy bands near the surface (band-bending). Such surface 2DEGs have provided the opportunities for direct investigations of their electronic structures using angle-resolved photoemission spectroscopy (ARPES). The ARPES studies of the 2DEGs at various surfaces have elucidated their fundamental characteristics \cite{Bi2Se3_2DEG, King_KTO100, Santander_Nature, Santander_KTO111} as well as their interplays with crystal symmetries \cite{Santander_KTO111, Wang_STO110, Hoesch_Bi2Te2Se}, spin-orbit coupling \cite{Sakano, Hoesch_Bi2Te2Se, King_Bi2Se3_Rashba, Santander_STO-gap}, and many-body interactions \cite{King_CdO, Barreto_EPC-2DEG}. Recently, intensive studies have been performed to tune the surface 2DEG carrier densities \cite{STO_2DEG, King_STO001, King_STO111, Rodel_TiO2, King_Bi2Se3_Rashba, Benia_Bi2Se3-tune, Bi2Se3_Suh}. In particular, at the surfaces of topological insulator Bi$_2$Se$_3$ and strongly correlated oxide SrTiO$_3$, modulation of the 2DEG densities is demonstrated by external means such as adsorption of atoms and photon irradiation \cite{STO_2DEG, King_STO001, King_STO111, Benia_Bi2Se3-tune}. Exploring further degrees of freedom in tunability of the surface 2DEGs by other means (e.g. by varying intrinsic physical parameters like local electric field and/or temperature) is of particular importance, since it may lead to the realization of exotic quantum phenomena and advanced devices utilizing the 2DEGs. Furthermore, it would be of considerable interest to find new materials possessing surface 2DEGs, where interplays between physical properties of the materials and the tunable surface 2DEGs can be explored.

In$_4$Se$_3$ is a quasi-one-dimensional semiconductor with the space group symmetry of $Pnnm$, which is represented by the corrugated layers of In-Se held together by the van der Waals force along the $a$-axis as shown in Fig. 1(a). While the quasi-two-dimensional electronic structure is expected from its weak bonding along the $a$-axis, the identification of nearly dispersionless bands along the $b$-axis classifies this material as quasi-one-dimensional\cite{FukutaniJPSJ}. Furthermore, In$_4$Se$_3$ is an excellent thermoelectric material characterized by large Seebeck coefficient and low thermal conductivity\cite{NatureTE}. 

In this paper, we report the observation of tunable 2DEG at the (100) surface of In$_4$Se$_3$ by means of high-resolution ARPES. The carrier density of the 2DEG exhibits a strong temperature dependence accompanied by a metal-semiconductor transition at the surface, which allows the thermal tuning of 2DEG at semiconductor surfaces. The present study suggests that the potentialities of 2DEGs can be further extended via controlling the intrinsic parameters of the materials. Moreover, utilization of such 2DEG on the state-of-the-art thermoelectric material may lead to another step forward in the enhancement of thermoelectric performance.

\section{EXPERIMENTS}
Single crystals of In$_4$Se$_3$ were grown by the Czochralski method. The deviation of the atomic composition from the stoichiometry (In:Se = 4:3) is estimated to be less than 1\% from the inductively coupled plasma (ICP) measurement. The high single crystallinity and the orientation of the samples were determined by the Laue diffraction measurements prior to the ARPES measurements. The samples were cleaved to expose the (100) surface, parallel to the \textit{bc}-plane as shown in Fig. 1(a). The cleaved surface shows mirror-like plane with only a few steps visible in optical microscope. The low energy electron diffraction (LEED) measurements have suggested no indication of surface reconstruction irrespective of temperature as shown in Fig. 1(c) and 1(d). We have also performed temperature-dependent electrical resistivity measurement for the bulk crystal to confirm the semiconducting nature of the sample as shown in Fig. 1(h). For each of the data presented, the different pieces of sample from the same parent crystal were freshly cleaved for the ARPES measurement, except for the series of photon-energy- and temperature-dependent measurements in Fig. 2(b), 2(c), 3(a) and 3(b), which were each performed upon same cleavage. The cleaving was performed at the temperature of 15 K. The ARPES measurements were performed in the ultra high vacuum (UHV) chamber at the base pressure of 3$\times$10$^{-11}$ torr using a VG-SCIENTA SES-2002 spectrometer with a high-flux helium discharge lamp and a toroidal grating monochromator. We used the He I$\alpha$, I$\beta$ and II$\alpha$ resonance lines (\textit{h}$\nu$ = 21.2, 23.1 and 40.8 eV, respectively) to excite photoelectrons. The energy and the angular resolutions were set to 40 meV and 0.2\degree, respectively. The Fermi level ($E_{\rm F}$) of the samples was referenced to that of a gold film evaporated onto the sample holder.

\begin{figure}
\includegraphics[width=3.6in]{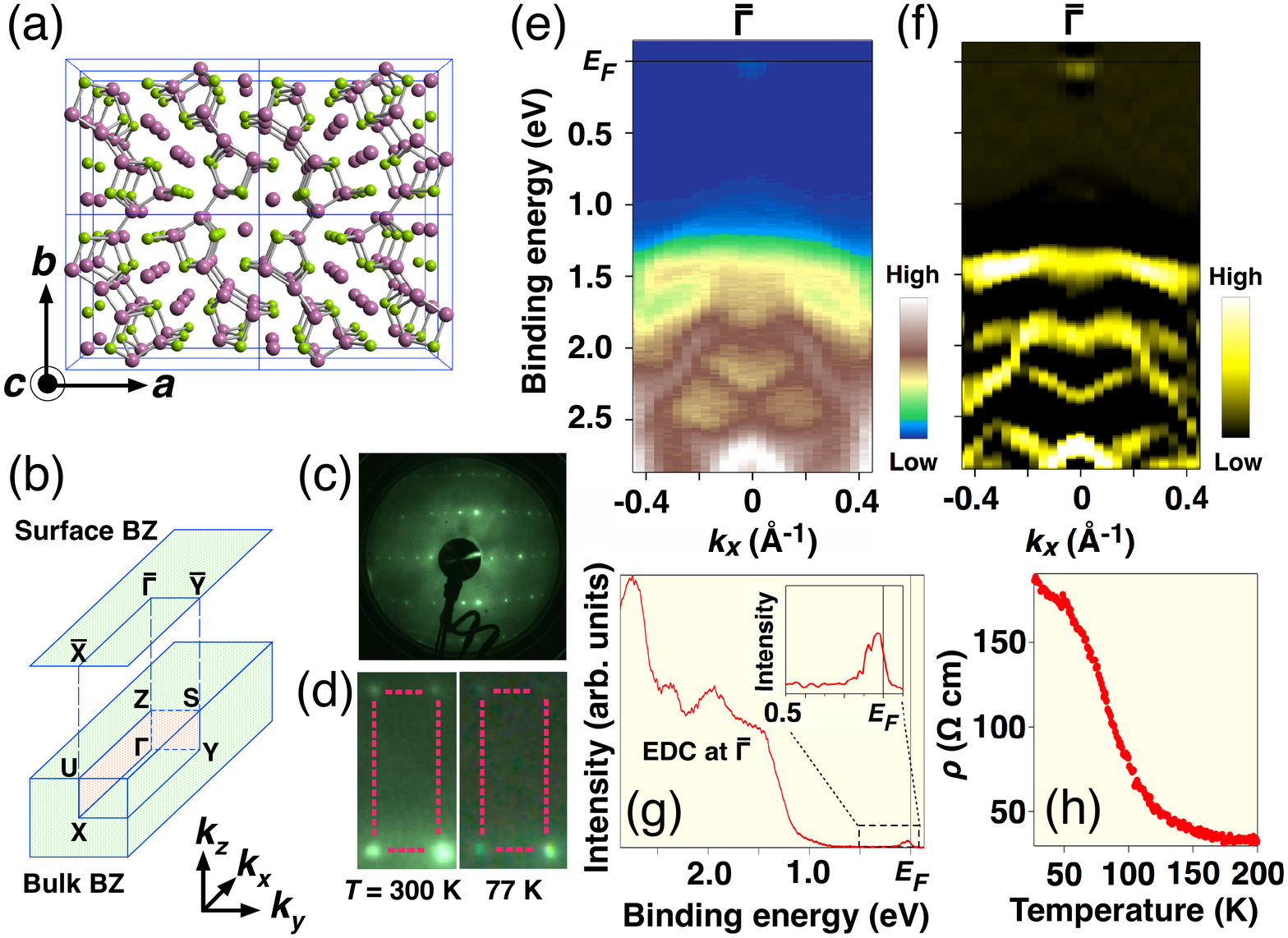}
\vspace{-0.5cm}
\caption{(Color online) (a) Crystal structure and (b) bulk and surface Brillouin zones of In$_4$Se$_3$. (c) The LEED image of In$_4$Se$_3$(100) surface observed at incident electron energy of 150 eV at $T$ = 300 K and (d) the close-up of LEED images at $T$ = 300 and 77 K, where the unit cell of the reciprocal lattice is marked by the dotted rectangles. (e) ARPES intensity and (f) corresponding second derivative plot in the valence-band region of In$_4$Se$_3$ measured along the $\bar{\Gamma}$-\={X} line with the He-I$\alpha$ ($h \nu$ = 21.2 eV) line at $T$ = 15 K. (g) Energy distribution curve (EDC) at the $\bar{\Gamma}$ point extracted from (e). Inset shows the magnified view of the near-$E_{\rm F}$ region indicated by dashed rectangle. (h) The electrical resistivity of In$_4$Se$_3$ as a function of temperature, exhibiting semiconducting behavior.}
\end{figure}

\begin{figure}
 \includegraphics[width=3.6in]{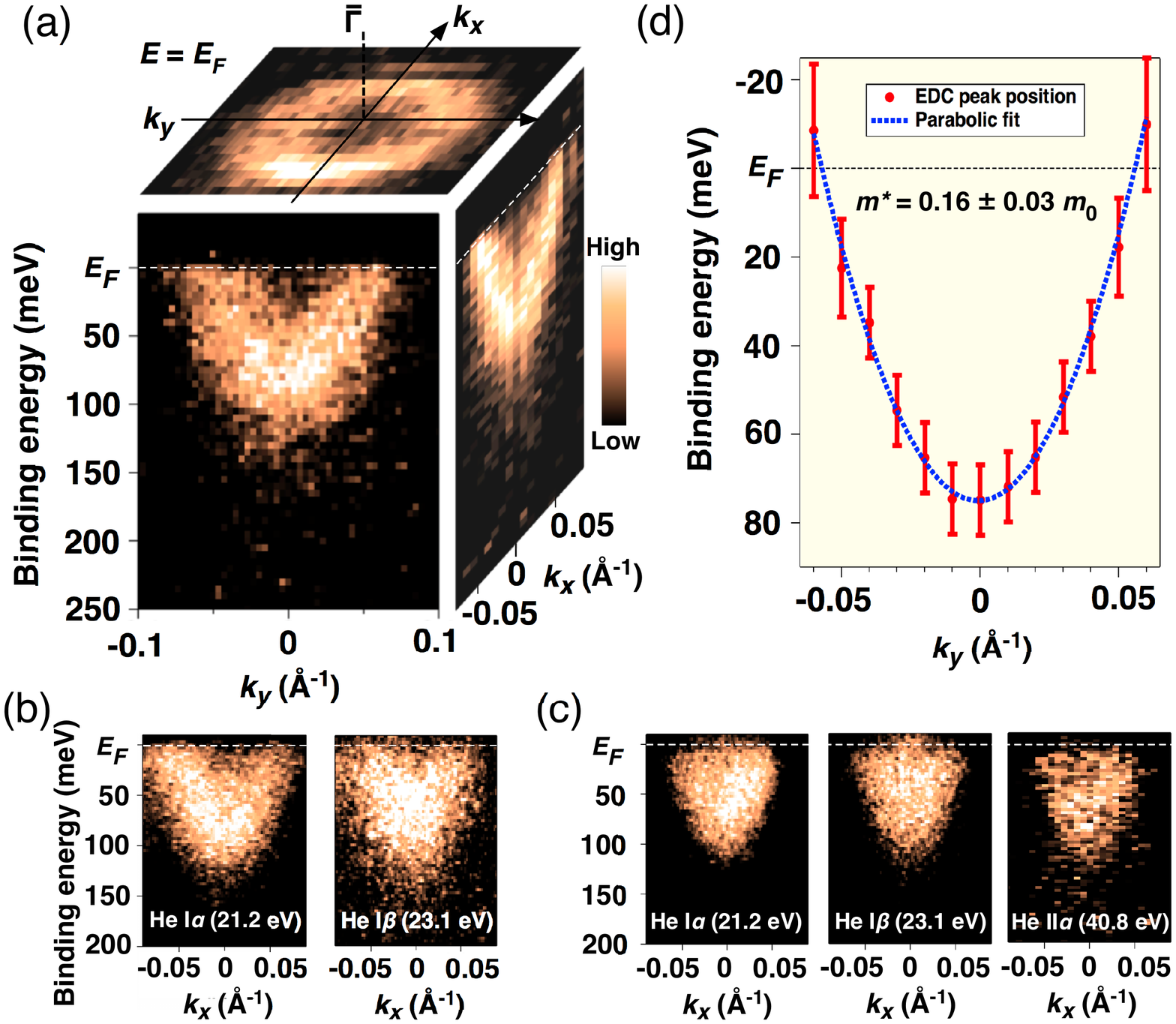}
 \vspace{-0.5cm}
 \caption{(Color online) (a) ARPES intensity plots of the 2DEG band at $T$ = 15 K. The top, front and side panels show the Fermi surface, the band dispersions along $\bar{\Gamma}$-\={Y} and $\bar{\Gamma}$-\={X} cuts, respectively. (b) Comparison of the ARPES intensity plots along the $\bar{\Gamma}$-\={X} cut at $T$ = 15 K measured with $h \nu$ = 21.2, 23.1 eV upon single cleavage. (c) Comparison of the ARPES intensity plots measured with $h \nu$ = 21.2, 23.1 and 40.8 eV upon single cleavage [different cleavage from (b) under otherwise same conditions]. (d) Peak positions of the energy band shown in (a) (red circles) extracted from the numerical fittings of the EDCs by Lorentzians multiplied by the Fermi-Dirac distribution function convoluted with the resolution function. The fitting of the experimental band dispersion with a parabolic function is also shown by dashed blue curve.}
\end{figure}

\section{RESULTS AND DISCUSSIONS}
Figures 1(e) and 1(f) show the ARPES-intensity plot and the corresponding second-derivative plot, respectively, for In$_4$Se$_3$ measured with the He-I$\alpha$ line along the $\bar{\Gamma}$-\={X} cut [\textit{k$_x$} cut; see Fig. 1(b)] at \textit{T} = 15 K in a relatively wide energy region with respect to $E_{\rm F}$. One can immediately recognize several dispersive bands below the binding energy of 1 eV, which originate from the bulk valence bands with mainly the Se 4\textit{p} orbital character \cite{FukutaniJPSJ}. A closer look in Figs. 1(f) and 1(g) reveals a finite intensity at around $E_{\rm F}$ at the $\bar{\Gamma}$ point, which is attributed to the bottom edge of the conduction band \cite{FukutaniJPSJ}. 

In order to identify this near-$E_{\rm F}$ state in more detail, we have performed ARPES experiments with higher energy/momentum resolutions and statistics. Figure 2(a) shows the ARPES-intensity plots along the \textit{k$_x$} ($\bar{\Gamma}$-\={X}) and \textit{k$_y$} ($\bar{\Gamma}$-\={Y}) cuts, together with the plot at $E_{\rm F}$ as a function of in-plane wave vectors \textit{k$_x$} and \textit{k$_y$}.  We clearly recognize a similar parabolic band in both directions, suggesting the circular shape of the Fermi surface. In fact, quantitative analysis indicates the Fermi wave vectors of $k_{\rm F}$ = 0.055$\pm$0.01 $\text{\AA}^{-1}$ and 0.055$\pm$0.005 $\text{\AA}^{-1}$ along the \textit{k$_x$} and \textit{k$_y$} axes, respectively, confirming the isotropic Fermi surface. 

It is crucial to identify whether this electronic state resides in the bulk or at the surface. For this purpose, we have performed ARPES measurements on two different cleavages at several photon energies, as shown in Fig. 2(b) and 2(c). First, it can be seen that depending on the surface cleavage, the binding energies of this band are slightly different [$\sim$70 meV for Fig. 2(b) and $\sim$40 meV for Fig. 2(c)], indicating the surface sensitivity of the state. This serves as a partial evidence for the two-dimensionality of the state as each cleavage was performed on samples taken from the same parent crystal (i.e., same bulk doping). Second, we note that for the same cleavage surface, it can be seen that the position of the band is nearly identical (within the experimental uncertainty of $\sim$20 meV) at the photon energies of 21.2 eV (He-I$\alpha$ line), 23.1 eV (He-I$\beta$) and 40.8 eV (He-II$\alpha$) as shown in Fig. 2(c) (the low intensity of the band at 40.8 eV is likely due to the photoemission matrix-element effect). While the unchanged position of the band at these different photon energies is consistent with the two-dimensionality for this state, it does not provide sufficient evidence here, because the particularly short $\Gamma$-Z distance in the bulk Brillouin zone along the $k_z$-direction could be covered by the $k_z$-broadening of probing region in photoemission \cite{FukutaniJPSJ} for all the photon energies utilized here. The third clarification for the dimensionality can be inferred from the absence of metallic feature in the temperature-dependent resistivity measurement shown in Fig. 1(h), which indicates that the metallic state observed in ARPES likely resides only near the surface (note that this metallic state observed in ARPES is present up to $T$$\sim$100 K, which is well covered by our resistivity measurement). The forth clarification of dimensionality of this state concerns with its temperature dependence. As will be discussed in detail below, the energy shift of this state caused by temperature variation exceeds the maximum possible shift set by the bulk Seebeck coefficient. This indicates that surface-related effect is necessary to account for the temperature-dependent behavior of this metallic state. The above set of evidences indicate that this near-$E_{\rm F}$ state most likely resides on the surface and is identified as 2DEG.

\begin{figure*}
 \includegraphics[width=6.4in]{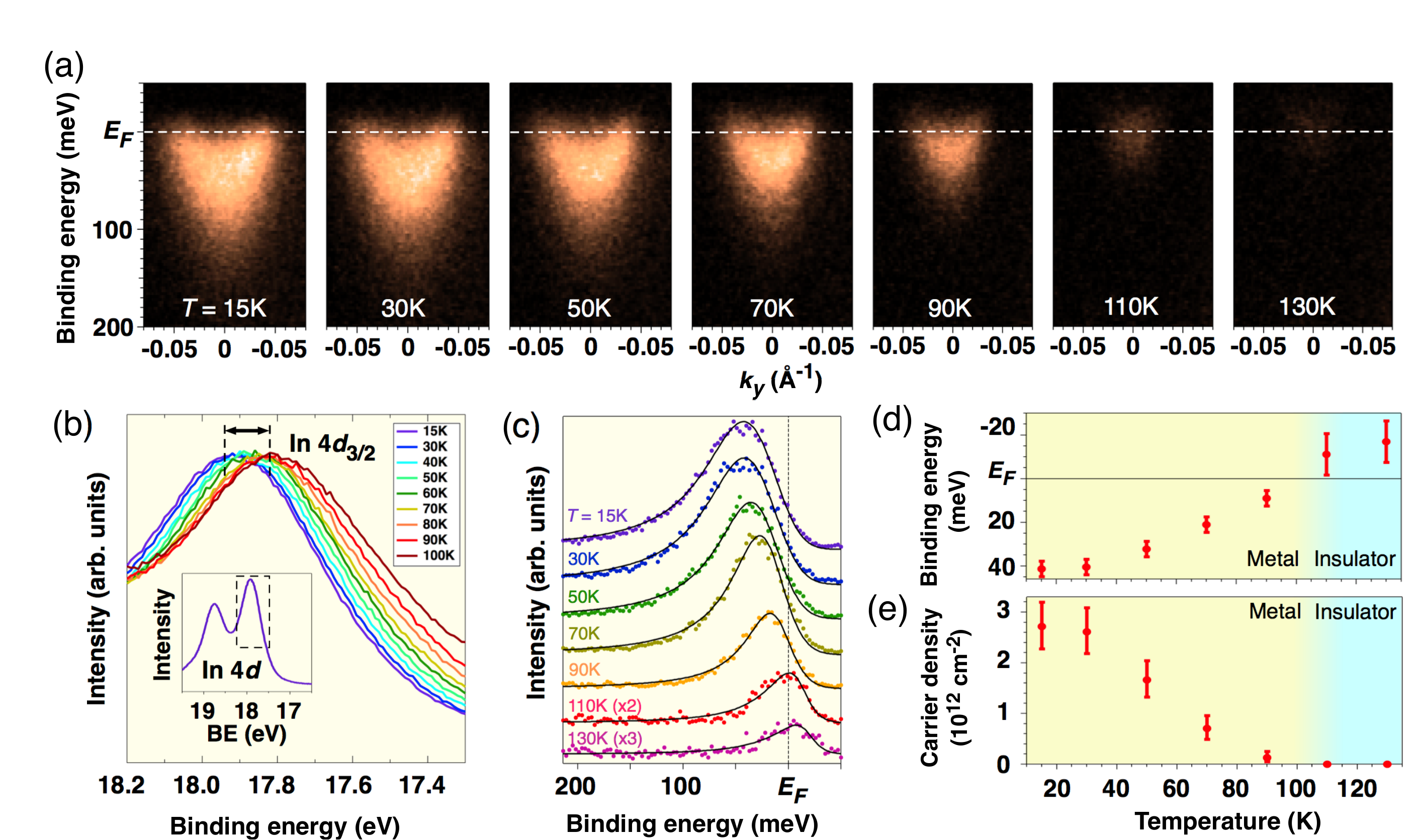}
 \vspace{-0.0cm}
 \caption{(Color online) (a) Temperature dependence of the ARPES intensity for the 2DEG band. (b) Temperature dependence of the In 4$d_{3/2}$ core level taken for the sample which exhibits the nearly identical band-bottom position at $T$ = 15 K as in (a). Inset shows the In 4$d$ core levels in a wider region (binding energy is abbreviated as BE). (c) Near-$E_{\rm F}$ EDCs at the $\bar{\Gamma}$ point for the data presented in (a) (dots) and the result of numerical fittings (solid curves). (d) Experimental position of the bottom of the 2DEG band as a function of temperature as determined from (c). (e) Estimated 2DEG density as a function of temperature.}
\end{figure*}

It is worthwhile to note that the observed 2DEG exhibits isotropic dispersion at the surface of quasi-one-dimensional crystal. A surface reconstruction into isotropic structure at low temperature can be excluded from the possible cause as can be seen from the LEED image in Fig. 1(d), which exhibits the strong anisotropy of the surface structure at $T$ = 77 K, where the isotropic 2DEG is observed. The possible explanation for the observation of isotropic state is that the bottom of the 2DEG band is too close to $E_{\rm F}$ to exhibit observable anisotropy. In fact, for the valence-band structure, our previous study shows that the topmost valence band is isotropic near the valence band maximum (VBM), while the quasi-one-dimensional bands were seen below $\sim$300 meV from the VBM \cite{FukutaniJPSJ}.

To extract essential physical parameters of the observed 2DEG, we have performed numerical fittings to the energy distribution curves (EDCs) with a Lorentzian multiplied by the Fermi-Dirac distribution function, convoluted with the experimental energy resolution. As displayed in Fig. 2(d), the extracted peak positions are well fitted by a parabolic function, indicating the free-electron-like character of the 2DEG. We have estimated the effective mass to be $m^*$ = 0.16$\pm$0.03 $m_0$ ($m_0$: free-electron mass), which is markedly smaller than those in oxides such as SrTiO$_3$ (Ref. \onlinecite{STO_2DEG}) and KTaO$_3$ (Ref. \onlinecite{King_KTO100}), but is similar to those in pnictides and other chalcogenides such as InAs (Ref. \onlinecite{King_CdO}) and Bi$_2$Se$_3$ (Ref. \onlinecite{Bi2Se3_2DEG}). This difference may arise from the character of atomic orbitals forming the 2DEGs, namely, delocalized $p$ orbitals (e.g. In 5$p$ and Bi 6$p$) would give rise to a smaller effective mass than that of localized $d$ orbitals (e.g. Ti 3$d$). It is noted that the small effective mass in In$_4$Se$_3$ is a great advantage in creating high mobility carriers.  

In order to explore the tunability of this 2DEG, we next measured the temperature dependence of the 2DEG band. Figure 3(a) shows the 2DEG band measured at various temperatures of 15-130 K upon same cleavage. The 2DEG band systematically moves upward upon increasing temperature. We have confirmed that this shift is reversible on a temperature cycle. At 110 and 130 K, we observe a very weak intensity at $E_{\rm F}$, which is associated with the tail of the peak located above $E_{\rm F}$. Such a systematic shift of the surface band is consistent with our earlier observation of temperature dependent shift of the valence band\cite{FukutaniJPSJ} and with the In 4$d$ core-level spectrum in Fig. 3(b), measured with the piece of sample which exhibits 2DEG band at nearly identical binding energy at $T$ = 15 K as in Fig. 3(a), which signifies a similar trend of the 4$d_{3/2}$ peak energy upon temperature variation [note that the larger shift in the core level ($\textgreater$100 meV for $T$ = 15-100 K) relative to that for the 2DEG band ($\sim$60 meV for $T$ = 15-130 K) is understood by the surface band-bending as discussed later].

We numerically fit the EDCs at the $\bar{\Gamma}$ point at each temperature by using the same procedures as in Fig. 2(c). The extracted energy positions of the bottom of the 2DEG band [Fig. 3(d)] shows that the total energy shift of the 2DEG band is $\sim$60 meV from 15 K to 130 K, and more importantly, at $T$ $\sim$ 100 K, the bottom of the 2DEG band crosses $E_{\rm F}$ in line with a sudden drop of ARPES intensity below $E_{\rm F}$ at 110 K. This indicates that there is no longer a metallic band at the surface above 100 K. We have estimated the carrier density of the 2DEG from the area of the circular Fermi surface, and the result is shown in Fig. 3(e). The carrier density exhibits an unconventional negative temperature dependence (i.e., it decreases with increasing temperature), in sharp contrast to the general properties observed in bulk semiconductors.  The carrier density is about 10$^{12}$ cm$^{-2}$ at 15-50 K, while it is almost zero at 100 K.  This strongly suggests that a temperature-induced metal-semiconductor transition takes place at the surface of In$_4$Se$_3$. It is worthwhile to point out that this significant variation in the carrier density and the metal-semiconductor transition is not detected by the electrical resistivity measurement as shown in Fig. 1(h), which further verifies that the observed electronic state resides only near the surface. It is also noted that the cleavage dependence of the 2DEG position would affect the exact transition temperature and carrier density, but the essential significant temperature dependence is independent from such quantitative variation.

Now we discuss the origin of the surface 2DEG in In$_4$Se$_3$. Since it is known that the 2DEG is formed to screen the excess positive charges of surface donor states \cite{Monch_book, Luth_book}, it is essential to reveal at first the origin of such positively charged (unoccupied) surface donor states. Since the LEED measurements [Fig. 1(c) and 1(d)] and ARPES results indicate no temperature-induced change in the band structure itself, the possibility of the surface reconstruction is excluded from the origin of positively charged surface states. Here we suggest number of possible origins. First, we consider effects of adsorbates on the surface. It is known that for Bi$_2$Se$_3$ adsorptions of residual gas molecules in the UHV chamber such as water \cite{Benia_Bi2Se3-tune} and CO (Ref. \onlinecite{CO-Bi2Se3}) are shown to induce strong band-bending, leading to the formation of 2DEG. Given the fact that Bi$_2$Se$_3$ and In$_4$Se$_3$ are both layered narrow-gap semiconductors containing Se, the effects of such adsorbates could play an important role in inducing the band-bending in In$_4$Se$_3$. On the other hand, we note that the 2DEG on In$_4$Se$_3$ is observed immediately after scanning over the entire Brillouin zone to determine $\bar{\Gamma}$ point (which takes less than 30 minutes after cleaving) and no subsequent shift of the 2DEG band is observed. Considering the pressure during the data acquisition (1$\times$10$^{-10}$ torr, where the rise from the base pressure is mostly due to He) and the adsorption time ($\textless$30 min.), the exposure to adsorbates is much less than 1 Langmuir, indicating that possible band-bending due to adsorbates saturates at very low coverage. This is in contrast to the surface 2DEG on Bi$_2$Se$_3$, where the band-bending continues to increase after $\textgreater$10 hours under similar experimental conditions \cite{CO-Bi2Se3}. Therefore, while the surface adsorbates could contribute to the band-bending, it is likely not a dominant effect. Second, we consider the effects pertaining to the cleaved surface, which arises from localized imperfections at the surface such as Se vacancies and excess In atoms. It is reported that for Bi$_2$Se$_3$, when Se is extracted from the surface, it leaves positively charged vacancy which in turn induces the downward surface band-bending \cite{Bi2Se3_2DEG, Bi2Se3_Suh}. Similarly for In$_4$Se$_3$, as the formal valence of Se is -2 (Ref. \onlinecite{LosovyjJAP}), the extraction of Se is expected to leave a positively charged vacancy resulting in the downward band-bending. Another possibility can be inferred from the crystal structure of In$_4$Se$_3$, in which two In atoms per unit cell are naturally intercalated (or ``sandwiched") within the Van der Waals gap between the atomic layers. While exactly one In atom per unit cell must be removed upon cleavage to maintain the charge neutrality, such a detailed balance is likely not achieved and there could be variations from the perfect charge neutrality (note that removal of all of these In atoms upon cleavage should not occur as it leaves the surface highly charged up). Thus, we also suggest some excess In atoms upon cleavage as one of the possible sources of positive charge at the surface. As the density of these defects would depend on the detailed surface conditions, the suggested possible origins pertaining to the cleaved surface for the surface 2DEG are at least consistent with the observed cleavage dependence of the band-bending strength, and should be investigated in the future studies by surface-sensitive microscopy methods such as scanning tunnelling microscopy (STM).

The origin of unusual temperature dependence of the 2DEG is also of considerable interest. There are mainly two possible contributions to the energy shift of the 2DEG; (i) temperature dependence of bulk chemical potential and (ii) temperature dependence of the band-bending strength. In order to estimate the first contribution, we note that the temperature-dependent chemical potential directly contributes to Seebeck coefficient \cite{Takeuchi}. Since the bulk Seebeck coefficient of In$_4$Se$_3$ has been reported to be \textbar$S$\textbar $\textless$ 200 $\mu$V/K for $T$ $\textless$ 130 K (Ref. \onlinecite{Shi}), it can be inferred that the upper limit for the temperature-induced shift of bulk chemical potential is $\sim$23 meV in the temperature range of 15-130 K (in which the Seebeck coefficient would be solely comprised of the chemical potential shift). Since this upper limit of the chemical-potential shift is much smaller than the observed shift of 2DEG ($\sim$ 60 meV), we conclude that the temperature dependence of the bulk chemical potential partially, but not fully, accounts for the temperature dependent shift of 2DEG (we note that this identification also implies that the observed electronic state should indeed be identified as surface 2DEG rather than a bulk state, as its energy shift in this temperature range exceeds the bulk limit of $\sim$23 meV). Nonetheless, as the sign of Seebeck coefficient is negative, the direction of shift is consistent with the observed shift of 2DEG and the large bulk Seebeck coefficient of this material likely plays a role in assisting the significant temperature dependence of the 2DEG. The second contribution must also play a crucial role, as the aforementioned bulk chemical-potential shift alone cannot fully account for the observed total energy shift of the 2DEG. Specifically, increasing temperature neutralizes the surface donor states by filling them with electrons via thermal excitations, thereby weakening the band-bending and \textit{vice versa}. While the energy of the core level states near the surface (measured by ARPES) directly follows such change in band-bending, the energy level of surface bound state (2DEG) generally moves much slower than the depth of the potential well \cite{King_CdO}. Such expected difference in the ``paces" of energy level shifts in the band-bending scenario is consistent with our observation that for $T$ = 15-100 K, the shifts of core level and 2DEG are more than 100 meV and $\sim$40 meV, respectively [see Figs. 3(b) vs. 3(d)]. It is therefore inferred that the donor-like surface state is located slightly above $E_{\rm F}$ in In$_4$Se$_3$ and the temperature-induced change in the band-bending strength is caused by their occupancy variations. The present study, however, does not identify any spectral evidence for this donor state, possibly due to its very weak intensity. With the proposed mechanisms of band-bending, the negative temperature dependence of 2DEG is not unique to In$_4$Se$_3$, and the underlying mechanism can be utilized to furnish other materials with a similar temperature tunability by appropriate surface doping to create the donor states slightly above $E_{\rm F}$. Furthermore, we remark that the temperature dependence of the band-bending here implies that the chemical potential at the surface shifts downward more rapidly than that of the bulk as a function of temperature. Thus, the Seebeck coefficient at the surface is likely larger than that of the bulk, consistent with the general expectation in reduced dimensionality.

The present finding of the surface 2DEG on In$_4$Se$_3$ and its significant temperature dependence show that an intrinsic degree of freedom can be utilized to control its carrier density and furthermore to switch the surface conductivity. The demonstration of such wide-range thermal tunability of the 2DEG likely serves to expand the future prospect of functional 2DEGs, as the carrier tunability of 2DEG is at the heart of modern electronics operations. The present study also expands the 2DEG material platform with excellent thermoelectric properties\cite{Bi2Se3_2DEG, TEandTopo}, where the interplay between the tunability of 2DEG and the large Seebeck coefficient can be utilized. The realization of 2DEG on In$_4$Se$_3$ implies that the utilization of such two-dimensional electrons could provide a promising platform to build high-performance thermoelectrics, for instance by enhancing the surface/volume ratio via the quantum-confinement effects through e.g., the fabrication of thin films and interfacing with other materials. In fact, such implementation has been demonstrated \cite{Dresselhaus_PbTe, Ohta_ZT, Ohta_Seebeck} for materials like SrTiO$_3$, and the addition of thermal tunability (negative temperature dependence) presented here could be used to overcome the general trend of decreasing Seebeck coefficient towards high temperature. Thus, when combined with the observed small effective mass ($\sim$0.16 $m_0$) of 2DEG and the exceptionally low thermal conductivity \cite{NatureTE, Shi, ThermalC_theory} of In$_4$Se$_3$, the tunable 2DEG on this material could be promising to achieve higher thermoelectric performance. In the future researches, it would be useful to unambiguously pin down the origin of the pertinent donor states and to reveal the band-bending profile by a detailed surface characterization and a comparison with theoretical models. Furthermore, to advance toward the device applications, identification of surface conductivity by surface sensitive probes on thin films\cite{Kim, SurfaceBi2Se3, Hasegawa} is much desired. 

\section{SUMMARY}
To summarize, we have identified the 2DEG at the (100) surface of In$_4$Se$_3$ by ARPES, and have demonstrated that the 2DEG exhibits strong temperature dependence, leading to metal-semiconductor transition at the surface. We attribute the significant temperature dependence to the high Seebeck coefficient of the material and the surface donor states lying close to $E_{\rm F}$.The present result serves to expand the degree of freedom in the controlling of 2DEG systems and could also provide additional ingredients to approach the improved thermoelectrics.

\begin{acknowledgements}
We thank P. A. Dowben for valuable discussions. This work was supported by Japan Society for the Promotion of Science (JSPS; KAKENHI 25287079, 15H02105), the Ministry of Education, Culture, Sports, Science and Technology of Japan (MEXT) (Innovative Area ``Topological Materials Science'', 15H05853, and "Science of Atomic Layers", 25107003), KEK-PF (Proposal No. 2015S2-003). 
\end{acknowledgements}

\newpage
\bibliographystyle{prsty}

\begin{thebibliography}{30}
\bibitem{Interface2DEG} R. Dingle, H. L. Stormer, and A.C. Gossard, Appl. Phys. Lett. \textbf{33}, 665 (1978). 
\bibitem{STO_LAO} A. Ohtomo and H. Y. Hwang, Nature (London) \textbf{427}, 423 (2004).
\bibitem{Klitzing} K. von Klitzing, G. Dorda, and M. Pepper, Phys. Rev. Lett. \textbf{45}, 494 (1980). 
\bibitem{Tsui} D. C. Tsui, H. L. Stormer, and A. C. Gossard, Phys. Rev. Lett. \textbf{48}, 1559 (1982). 
\bibitem{2Dsuper} N. Reyren, S. Thiel, A. D. Caviglia, L. F. Kourkoutis, G. Hammerl, C. Richter, C. W. Schneider, T. Kopp, A. S. R\"{u}etschi, D. Jaccard, M. Gabay, D. A. Muller, J. M. Triscone, and J. Mannhart, Science \textbf{317}, 1196 (2007).
\bibitem{RMP_Ando} T. Ando, A. B. Fowler, and F. Stern, Rev. Mod. Phys. \textbf{54}, 437 (1982). 
\bibitem{Dresselhaus_PbTe} L. D. Hicks, T. C. Harman, X. Sun, and M. S. Dresselhaus, Phys. Rev. B \textbf{53}, 66707 (1996).
\bibitem{Ohta_ZT} H. Ohta, S. Kim, Y. Mune, T. Mizoguchi, K. Nomura, S. Ohta, T. Nomura, Y. Nakanishi, Y. Ikuhara, M. Hirano, H. Hosono, and K. Koumoto, Nat. Mater. \textbf{6}, 129 (2007).
\bibitem{Ohta_Seebeck} H. Ohta, T. Mizuno, S. Zheng, T. Kato, Y. Ikuhara, K. Abe, H. Kumomi, K. Nomura, and H. Hosono, Adv. Mater. \textbf{24}, 740 (2012).
\bibitem{SiGe} T. Koga, S. B. Cronin, M. S. Dresselhaus, J. L. Liu and K. L. Wang, Appl. Phys. Lett. \textbf{77}, 1490 (2000).
\bibitem{Bi2Te3} R. Venkatasubramanian, E. Siivola, T. Colpitts and B. O'Quinn, Nature (London) \textbf{413}, 597 (2001).
\bibitem{Harman} T. C. Harman, M. P. Walsh, B. E. LaForge and G. W. Turner, J. Electron. Mater. \textbf{34}, L19 (2005).
\bibitem{ReviewTE} M. S. Dresselhaus, G. Chen, M. Y. Tang, R. Yang, H. Lee, D. Wang, Z. Ren, J.-P. Fleurial and P. Gogna, Adv. Mater. \textbf{19}, 1043 (2007).
\bibitem{Bi2Se3_2DEG} M. Bianchi, D. Guan, S. Bao, J. Mi, B. B. Iversen, P. D. C. King, and P. Hofmann, Nat. Commun. \textbf{1}, 128 (2010).
\bibitem{King_KTO100} P. D. C. King, R. H. He, T. Eknapakul, P. Buaphet, S. K. Mo, Y. Kaneko, S. Harashima, Y. Hikita, M. S. Bahramy, C. Bell, Z. Hussain, Y. Tokura, Z. X. Shen, H. Y. Hwang, F. Baumberger, and W. Meevasana, Phys. Rev. Lett. \textbf{108}, 117602 (2012).
\bibitem{Santander_Nature} A. F. Santander-Syro, O. Copie, T. Kondo, F. Fortuna, S. Pailh\`{e}s, R. Weht, X. G. Qiu, F. Bertran, A. Nicolaou, A. Taleb-Ibrahimi, P. Le F\`{e}vre, G. Herranz, M. Bibes, N. Reyren, Y. Apertet, P. Lecoeur, A. Barth\'{e}l\'{e}my, and M. J. Rozenberg, Nature (London) \textbf{469}, 189 (2011).
\bibitem{Santander_KTO111} C. Bareille, F. Fortuna, T. C. Rodel, F. Bertran, M. Gabay, O. H. Cubelos, A. Taleb-Ibrahimi, P. Le F\`{e}vre, M. Bibes, A. Barth\'{e}l\'{e}my, T. Maroutian, P. Lecoeur, M. J. Rozenberg, and A. F. Santander-Syro, Sci. Rep. \textbf{4}, 3586 (2014).
\bibitem{Wang_STO110} Z. Wang, Z. Zhong, X. Hao, S. Gerhold, B. Stoeger, M. Schmid, J. Sanchez-Barriga, A. Varykhalov, C. Franchini, K. Held, and U. Diebold, Proc. Natl. Acad. Sci. USA \textbf{111}, 3933 (2014).
\bibitem{Hoesch_Bi2Te2Se} M. Michiardi, M. Bianchi, M. Dendzik, J. A. Miwa, M. Hoesch, T. K. Kim, P. Matzen, J. Mi, M. Bremholm, B. B. Iversen, and P. Hofmann, Phys. Rev. B \textbf{91}, 035445 (2015).
\bibitem{Sakano} M. Sakano, M. Bahramy, A. Katayama, T. Shimojima, H. Murakawa, Y. Kaneko, W. Malaeb, S. Shin, K. Ono, H. Kumigashira, R. Arita, N. Nagaosa, H. Hwang, Y. Tokura, and K. Ishizaka, Phys. Rev. Lett. \textbf{110}, 107204 (2013).
\bibitem{King_Bi2Se3_Rashba} P. D. C. King, R. C. Hatch, M. Bianchi, R. Ovsyannikov, C. Lupulescu, G. Landolt, B. Slomski, J. H. Dil, D. Guan, J. L. Mi, E. D. L. Rienks, J. Fink, A. Lindblad, S. Svensson, S. Bao, G. Balakrishnan, B. B. Iversen, J. Osterwalder, W. Eberhardt, F. Baumberger, and P. Hofmann, Phys. Rev. Lett. \textbf{107}, 096802 (2011).
\bibitem{Santander_STO-gap} A. F. Santander-Syro, F. Fortuna, C. Bareille, T. C. Rodel, G. Landolt, N. C. Plumb, J. H. Dil, and M. Radovi\'{c}, Nat. Mater. \textbf{13}, 1085 (2014).
\bibitem{King_CdO} P. D. C. King, T. D. Veal, C. F. McConville, J. Z\'{u}\~{n}iga-P\'{e}rez, V. Mu\~{n}oz-Sanjos\'{e}, M. Hopkinson, E. D. L. Rienks, M. F. Jensen, and P. Hofmann, Phys. Rev. Lett. \textbf{104}, 256803 (2010).
\bibitem{Barreto_EPC-2DEG} L. Barreto, M. Bianchi, D. Guan, R. Hatch, J. Mi, B. B. Iversen, and P. Hofmann, Phys. Status Solidi (RRL) \textbf{7}, 136 (2012).
\bibitem{STO_2DEG} W. Meevasana, P. D. C. King, R. H. He, S. K. Mo, M. Hashimoto, A. Tamai, P. Songsiriritthigul, F. Baumberger, and Z. X. Shen, Nat. Mater. \textbf{10}, 114 (2011).
\bibitem{King_STO001} S. M. Walker, F. Y. Bruno, Z. Wang, A. de la Torre, S. Ricc\'{o}, A. Tamai, T. K. Kim, M. Hoesch, M. Shi, M. S. Bahramy, P. D. C. King, and F. Baumberger, Adv. Mater. \textbf{27}, 3894 (2015).
\bibitem{King_STO111} S. McKeown Walker, A. de la Torre, F. Y. Bruno, A. Tamai, T. K. Kim, M. Hoesch, M. Shi, M. S. Bahramy, P. D. C. King, and F. Baumberger, Phys. Rev. Lett. \textbf{113}, 177601 (2014).
\bibitem{Rodel_TiO2} T. C. Rodel, F. Fortuna, F. Bertran, M. Gabay, M. J. Rozenberg, A. F. Santander-Syro, and P. Le F\`{e}vre, Phys. Rev. B \textbf{92}, 041106 (2015).
\bibitem{Benia_Bi2Se3-tune} H. M. Benia, C. Lin, K. Kern, and C. R. Ast, Phys. Rev. Lett. \textbf{107}, 177602 (2011).
\bibitem{Bi2Se3_Suh}	J. Suh, D. Fu, X. Liu, J. K. Furdyna, K. M. Yu, W. Walukiewicz, and J. Wu, Phys. Rev. B \textbf{89}, 115307 (2014).
\bibitem{FukutaniJPSJ} K. Fukutani, Y. Miyata, I. Matsuzaki, P. V. Galiy, P. A. Dowben, T. Sato, T. Takahashi, J. Phys. Soc. Jpn. \textbf{84}, 074710 (2015).
\bibitem{NatureTE}	J.-S. Rhyee, K. H. Lee, S. M. Lee, E. Cho, S. Il Kim, E. Lee, Y. S. Kwon, J. H. Shim, and G. Kotliar, Nature \textbf{459}, 965 (2009).
\bibitem{Monch_book}W. M\"{o}nch, \textit{Semiconductor surfaces and interfaces} (Springer, Berlin, 2001).
\bibitem{Luth_book} H. Luth, \textit{Solid surfaces, interfaces and thin films} (Springer, Berlin, 2010).
\bibitem{CO-Bi2Se3} M. Bianchi, R. C. Hatch, D. Guan, T. Planke, J. Mi, B. B. Iversen, and P. Hoffmann, Semicond. Sci. Technol. \textbf{27}, 124001 (2012).
\bibitem{LosovyjJAP} Y. B. Losovyj, L. Makinistian, E. A. Albanesi, A. G. Petukhov, J. Liu, P. Galiy, O. R. Dveriy, and P. A. Dowben, J. Appl. Phys. \textbf{104}, 083713 (2008).
\bibitem{Takeuchi} T. Takeuchi, Y. Toyama and A. Yamamoto, Mater. Trans. \textbf{51}, 421 (2010).
\bibitem{Shi}X. Shi, J. Y. Cho, J. R. Salvador, J. Yang, and H. Wang, Appl. Phys. Lett. \textbf{96}, 162108 (2010).
\bibitem{TEandTopo} H. Osterhage, J. Gooth, B. Hamdou, P. Gwozdz, R. Zierold and K. Nielsch, Appl. Phys. Lett. \textbf{105}, 123117 (2014).
\bibitem{ThermalC_theory} H. S. Ji, H. Kim, C. Lee, J.-S. Rhyee, M. H. Kim, M. Kaviany, and J. H. Shim, Phys. Rev. B \textbf{87}, 125111 (2013).
\bibitem{Kim} D. Kim, S. Cho,	N. P. Butch, P. Syers, K. Kirshenbaum, S. Adam, J. Paglione and M. S. Fuhrer, Nature Phys. \textbf{8}, 459 (2012).
\bibitem{SurfaceBi2Se3} J. G. Checkelsky, Y. S. Hor, R. J. Cava, and N. P. Ong, Phys. Rev. Lett. \textbf{106}, 196801 (2011).
\bibitem{Hasegawa} S. Hasegawa, I. Shiraki, F. Tanabe and R. Hobara, Curr. Appl. Phys. \textbf{2}, 465 (2002).
\end{thebibliography}

\end{document}